\documentclass{prop2015}
\usepackage[english]{babel}
\usepackage{amsmath}
\usepackage{cuted}

\newcommand{\SLf}{\mathrm{SL}(5)}
\newcommand{\SLfo}{\mathrm{SL}(4)}
\newcommand{\bba}{\bar{a}}
\newcommand{\bbb}{\bar{b}}
\newcommand{\bbc}{\bar{c}}
\newcommand{\bbd}{\bar{d}}
\newcommand{\bee}{\bar{e}}
\newcommand{\bbf}{\bar{f}}

\newcommand{\bfi}{\bar{5}}

\newcommand{\balpha}{\bar{\alpha}}
\newcommand{\bbeta}{\bar{\beta}}
\newcommand{\bgamma}{\bar{\gamma}}
\newcommand{\bdelta}{\bar{\delta}}

        {%
            \par  \vspace{-5pt}
            \hfill\rule[-6pt]{0.4pt}{6.4pt}%
            \rule{\dimexpr(0.5\textwidth-0.5\columnsep-1pt)}{0.4pt}
            \end{strip}
        }

\title{Dualising consistent truncations}
\author[E. Malek]{Emanuel Malek\inst{1}\footnote{E-mail:~\textsf{E.Malek@lmu.de}}}
\address[1]{Arnold Sommerfeld Center for Theoretical Physics,\\ Ludwig-Maximilians-Universit\"at M\"unchen, Theresienstra{\ss}e 37, 80333 M\"unchen, Germany.}

\begin{abstract}
 We use exceptional field theory to establish a duality between certain consistent 7-dimensional truncations with maximal SUSY from IIA to IIB. We use this technique to obtain new consistent truncations of IIB on $S^3$ and $H^{p,q}$ and also give a no-go theorem for the uplift of certain gaugings.
\end{abstract}
\shortabstract

\begin{acknowledgements}
 The author would like to thank the organisers of ``The String Theory Universe, 21st European Workshop and 3rd COST MP1210 meeting'' for the opportunity to present this work and would like to thank Henning Samtleben for the collaboration on which this talk is based. The author is supported by the ERC Advanced Grant ``Strings and Gravity" (Grant No. 32004) and was also funded by the National Research Foundation (NRF) of South Africa under grant CSUR13091742207 during the initial stages of this project.
\end{acknowledgements}

\begin{document}
\maketitle

\section{Introduction}
The consistent Kaluza-Klein truncation of higher-di\-men\-sion\-al supergravity to lower-dimensional theories is an old and generically difficult problem due to the highly non-linear gravitational field equations~\cite{Duff:1984hn}. In the case of truncations on group manifolds via a Scherk-Schwarz Ansatz, consistency follows from the group structure. However, consistent reductions can also be obtained without compactifying on a group manifold, as is the case for the sphere reductions of 11-dimensional SUGRA~\cite{deWit:1986iy,Nastase:1999kf,Cvetic:2000dm}. Recent progress in understanding these consistent truncations has come via generalised Scherk-Schwarz reductions~\cite{Aldazabal:2011nj,Geissbuhler:2011mx,Berman:2012uy,Hohm:2014qga,Lee:2014mla,Lee:2015xga} in double field theory (DFT)~\cite{Hull:2009mi} and exceptional field theory (EFT)~\cite{Berman:2010is}, and their associated generalised geometries~\cite{Coimbra:2011ky}.

In \cite{Malek:2015hma} we use this framework to establish a duality relating consistent IIA and IIB truncations for certain gaugings of maximal 7-dimensional supergravity. We then employ this duality to derive new consistent truncations of type IIB theory on the three sphere $S^3$, as well as on hyperboloids $H^{p,q}$, which lead to compact $SO(4)$, non-compact $SO(p,q)$ and non-semisimple $CSO(p,q,r)$ gaugings.

\section{Seven-dimensional maximal gauged SUGRA}
Maximal seven-dimensional gauged SUGRA can be conveniently formulated using the embedding tensor formalism \cite{Samtleben:2005bp}, which describes the embedding of the gauge group into the global symmetry group of the ungauged theory, in this case $\SLf$. Let us denote fundamental $\SLf$ indices by $\bba, \bbb = 1, \ldots 5$ and observe that the gauge group can at most be 10-dimensional because there are only 10 vector fields in seven-dimensional maximal SUGRA. If we denote by $X_{\bba\bbb}$ the 10 generators of the gauge group and $t_\alpha$ the generators of $\SLf$, we have
\begin{equation}
 X_{\bba\bbb} = \theta_{\bba\bbb}{}^\alpha t_\alpha \,,
\end{equation}
where $\alpha$ labels the adjoint of $\SLf$ and $\theta_{\bba\bbb}{}^\alpha$ is the embedding tensor.

Consistency of the seven-dimensional theory requires the embedding tensor to satisfy two constraints. The linear constraint restricts the representations in which the embedding tensor must lie:
\begin{equation}
 \theta_{\bba\bbb}{}^\alpha \in \mathbf{15} \oplus \mathbf{40}' \oplus \mathbf{10} \subset \mathbf{10} \otimes \mathbf{24} \,,
\end{equation}
and ensures supersymmetry. We will denote these irreps by $S_{\bba\bbb}$, $Z^{\bba\bbb,\bbc}$ and $\tau_{\bba\bbb}$ for the $\mathbf{15}$, $\mathbf{40}'$ and $\mathbf{10}$, respectively. Let us note that gaugings in the $\mathbf{10}$, known as ``trombone gaugings'', lead to gauged SUGRAs which can only be defined at the level of the equations of motion since they do not admit an action principle~\cite{LeDiffon:2008sh}. The quadratic constraint ensures closure of the gauge group
\begin{equation}
 \left[ X_{\bba\bbb}, X_{\bbc\bbd} \right] = - \left(X_{\bba\bbb,\bbc\bbd}\right)^{\bee\bbf} X_{\bee\bbf} \,, \label{eq:QC}
\end{equation}
where
\begin{equation}
 \left(X_{\bba\bbb}\right)_{\bbc}{}^{\bbd} = \frac{1}{2} \epsilon_{\bba\bbb\bbc\bee\bbf} Z^{\bee\bbf,\bbd} + 2 \delta^{\bbd}_{[\bba} S_{\bbb]\bbc} + \frac{1}{3} \delta_{\bbc}^{\bbd} \tau_{\bba\bbb} + \frac{2}{3} \delta^{\bbd}_{[\bba} \tau_{\bbb]\bbc} \,,
\end{equation}
and $\epsilon_{\bba\bbb\bbc\bbd\bee} = \epsilon^{\bba\bbb\bbc\bbd\bee} = \pm 1$ is the 5-d alternating symbol. Here we will primarily be interested in the scalar sector of the gauged SUGRA, which parameterises the coset space $\SLf / \mathrm{SO}(5)$ and can thus be represented by a unit-determinant $5\times 5$ matrix $M_{\bba\bbb}$. In addition, the theory also contains 10 vectors $A_\mu{}^{\bba\bbb}$, five two-forms $B_{\mu\nu\,\bba}$, five three-forms $C_{\mu\nu\rho}{}^{\bba}$ and the seven-dimensional metric $G_{\mu\nu}$.

Interestingly, the quadratic constraint \eqref{eq:QC} allows for two inequivalent types of $\textrm{CSO}(p,q,r)$ gaugings \cite{Samtleben:2005bp}. The first is triggered by gaugings in the $\mathbf{15}$ and is obtained by truncating M-theory on the hyperboloid $H^{p,q} \times T^r$, while the second is triggered by certain gaugings in the $\mathbf{40}'$ and is believed to be related to truncations of IIB supergravity on $H^{p,q} \times T^r$. We will now use EFT \cite{Berman:2010is} to show that these IIA / IIB truncations can be dualised into each other.

\section{Generalised Scherk-Schwarz truncations}
Our key tool for studying seven-dimensional consistent truncations is the $\SLf$ EFT \cite{Berman:2010is}. This is a reformulation of 11- and 10-dimensional supergravity which renders manifest the $\SLf$ symmetry known to appear under dimensional reduction to seven dimensions. In order to do this, extra coordinates are introduced so that there are 10 ``internal coordinates'' $Y^{ab}$ transforming in the antisymmetric representation of $\SLf$. In the case of a toroidal compactification of 11-dimensional supergravity, the extra six coordinates can be understood as duals to the wrapping modes of membranes.

With respect to the seven-dimensional ``external'' spacetime, the degrees of freedom of 11- and 10-dimensional supergravity coincide with those of the maximal gauged SUGRA and we denote them by the same symbols in calligraphic font: ${\cal M}_{ab}$, ${\cal A}_\mu{}^{ab}$, ${\cal B}_{\mu\nu\,a}$ and ${\cal C}_{\mu\nu\rho}{}^a$, ${\cal G}_{\mu\nu}$. These fields can depend on all of the external coordinates, $x^\mu$, but their dependence on the internal coordinates, $Y^{ab}$, is restricted by the so-called ``section condition'' \cite{Berman:2011cg}
\begin{equation}
 \partial_{[ab} f(x,Y)\, \partial_{cd]} g(x,Y) = \partial_{[ab} \partial_{cd]} f(x,Y) = 0 \,,
\end{equation}
for all physical fields which we symbolically denote here by $f(x,Y)$ and $g(x,Y)$. The section condition can be solved by only allowing dependence on four coordinates $Y^{i5}$ where $i = 1, \ldots 4$ upon which the theory reduces to the full 11-dimensional supergravity.

Here we wish to consider the type II theories and so it is useful to decompose $\SLf \rightarrow \SLfo \simeq \mathrm{Spin}(3,3)$, the relevant ``T-duality'' group. The coordinate representation then splits as $\mathbf{10} \rightarrow \mathbf{6} \oplus \mathbf{4}$ where the $\mathbf{6}$ consists of the 3 coordinates of IIA plus the 3 coordinates of IIB. Thus to consider only the type II theories, we restrict the coordinate dependence to the $\mathbf{6}$ by demanding all fields to be independent of $Y^{\alpha 5}$, where $\alpha = 1, \ldots 4$. Now the section condition reduces to
\begin{equation}
 \partial_{\alpha\beta} f(x,Y) \partial^{\alpha\beta} g(x,Y) = \partial_{\alpha\beta} \partial^{\alpha\beta} f(x,Y) = 0 \,,
\end{equation}
where we define $\partial^{\alpha\beta} = \frac{1}{2} \epsilon^{\alpha\beta\gamma\delta} \partial_{\gamma\delta}$. There are two inequivalent solutions to this constraint, where all fields only depend on either
\begin{equation}
 y^m \equiv Y^{m4} ~~ (\mathrm{IIA}) \,, \qquad \mathrm{or } \qquad  \tilde{y}_m \equiv \frac{1}{2} \epsilon_{mnp} Y^{np}  ~~ (\mathrm{IIB}) \,,
\end{equation}
with $m = 1, \ldots 3.$ Upon restricting the field dependence as above, the EFT reduces to either IIA or IIB SUGRA, as has been shown for the scalar sector in \cite{Blair:2013gqa}, and for the full EFT in other dimensions, for example in \cite{Hohm:2013uia} for the $E_{7(7)}$ EFT. An important observation for what follows is that these two solutions of the section condition are related by the $\mathbb{Z}_2$ outer automorphism of $\SLfo$ which takes \\$\partial_{\alpha\beta} \longleftrightarrow \partial^{\alpha\beta}$.

Given this framework, a natural way to generalise the Scherk-Schwarz Ansatz is to allow for a $\SLf$-valued twist $U_a{}^{\bba}(Y)$ of the EFT fields, as follows \cite{Hohm:2014qga}
\begin{equation}
 \begin{split}
  \mathcal{M}_{ab}\left(x,Y\right) &= U_{a}{}^{\bba}(Y)\,U_{b}{}^{\bbb}(Y)\,{M}_{\bba\bbb}(x)\,, \\
  G_{\mu\nu}(x,Y) &= \rho^{-2}(Y)\,G_{\mu\nu}(x)\,,\\
  {\cal A}_{\mu}{}^{ab}(x,Y) &= \rho^{-1}(Y)\, A_{\mu}{}^{\bar{a}\bar{b}}(x)U_{\bar{a}\bar{b}}{}^{ab}(Y) \,, \\
  {\cal B}_{\mu\nu\,a}(x,Y) &= \rho^{-2}(Y)\, B_{\mu\nu\,\bar{a}}(x)\,U_{a}{}^{\bar{a}}(Y) \,, \\
  {\cal C}_{\mu\nu\rho}{}^a(x,Y) &= \rho^{-3}(Y)\, C_{\mu\nu\rho}{}^{\bar{a}}(x)\,U_{\bar{a}}{}^{a}(Y) \,,
  \label{eq:SSAnsatz}
 \end{split}
\end{equation}
where $\rho(Y)$ is some scalar function and $U_{\bba\bbb}{}^{ab} = U_{\bba}{}^{[a} U_{\bbb}{}^{b]}$. The metric and p-form fields in the internal directions can be read off from ${\cal M}_{ab}$, with the explicit formulae given in \cite{Malek:2015hma}, while the remaining degrees of freedom of the 10- or 11-dimensional supergravity are encoded in ${\cal A}_{\mu}{}^{ab}$, ${\cal B}_{\mu\nu\,a}$ and ${\cal C}_{\mu\nu\rho}{}^a$.

This Ansatz leads to a maximal gauged SUGRA with the embedding tensor as a function of the twists~\cite{Berman:2012uy}, which must be nowhere vanishing. In particular, the embedding tensor is realised as the torsion of the flat connection of the extended space \cite{Coimbra:2011ky,Berman:2013uda,Blair:2014zba}
\begin{equation}
 \begin{split}
  S_{\bar{a}\bar{b}} &= \frac{1}{\rho} \partial_{ab} U_{(\bar{a}}{}^a U_{\bar{b})}{}^{b} \,, \\
  Z^{\bar{a}\bar{b},\bar{c}} &= \frac{1}{2\rho} \epsilon^{abcef} \left( U_{ef}{}^{\bar{a}\bar{b}} \, \partial_{ab} U_c{}^{\bar{c}} - U_{ef}{}^{[\bar{a}\bar{b}}\,\partial_{ab} U_c{}^{\bar{c}]} \right) \,, \\
  \tau_{\bar{a}\bar{b}} &= - \frac{1}{2\rho}\left( \partial_{cd} U_{\bar{a}\bar{b}}{}^{cd} - 6\,\rho^{-1} \, U_{\bar{a}\bar{b}}{}^{cd} \,\partial_{cd} \rho \right) \,.
 \end{split}
 \label{eq:consistent}
\end{equation}
One can show that the reduction is consistent if the embedding tensor is constant~\cite{Aldazabal:2011nj,Geissbuhler:2011mx,Berman:2012uy}, thus providing a set of differential equations which have to be solved for the twists $U_a{}^{\bba}(Y)$ and $\rho(Y)$. A sufficient requirement for the embedding tensor to satisfy the quadratic constraint is that the twists obey the section condition.

\section{Dualising consistent truncations}
Having reviewed the set-up for generalised Scherk-Schwarz truncations, let us now turn to the question of whether and when truncations of IIA have dual truncations of IIB and vice versa. To do so, let us decompose the embedding tensor under $\SLf \rightarrow \SLfo \simeq \mathrm{Spin}(3,3)$ to find
\begin{equation}
 \begin{split}
  \mathbf{15} &\rightarrow \mathbf{10} \oplus \mathbf{4} \oplus \mathbf{1} \,, \qquad \mathbf{10}' \rightarrow \mathbf{6} \oplus \mathbf{4} \,, \\
  \mathbf{40}' &\rightarrow \mathbf{20}' \oplus \mathbf{10}' \oplus \mathbf{6} \oplus \mathbf{4}' \,.
 \end{split}
\end{equation}

Let us for now make a $\mathrm{GL}(4)$ Ansatz for the twist matrix
\begin{equation}
 U_a{}^{\bba} = \begin{pmatrix}
  \omega^{-1/2} V_\alpha{}^{\balpha} & 0 \\ 0 & \omega^2
 \end{pmatrix} \,, \qquad |V| = 1 \,. \label{eq:DiagAnsatz}
\end{equation}
From \eqref{eq:consistent} one finds that this only generates non-zero gaugings in the $\mathbf{10}$'s and $\mathbf{6}$'s:
\begin{equation}
 \begin{split}
  S_{\balpha\bbeta} &\equiv M_{\balpha\bbeta} = \rho^{-1} \omega V_{(\balpha}{}^\alpha \partial_{|\alpha\beta|} V_{\bbeta)}{}^\beta \,, \\
  Z^{\bfi(\balpha,\bbeta)} &\equiv \tilde{M}^{\balpha\bbeta} = \rho^{-1} \omega V_\alpha{}^{(\balpha} \partial^{|\alpha\beta|} V_\beta{}^{\bbeta)} \,, \\
  2 \tau_{\balpha\bbeta} &= - \rho^{-1} \omega \left( \partial_{\alpha\beta} V_{\balpha\bbeta}{}^{\alpha\beta} -5 V_{\balpha\bbeta}{}^{\alpha\beta} \partial_{\alpha\beta} \ln \omega \right. \\
  & \left. \quad + \,6 V_{\balpha\bbeta}{}^{\alpha\beta} \partial_{\alpha\beta} \ln \left(\rho^{-1}\omega\right) \right)\,, \\
  6Z^{\bfi[\balpha,\bbeta]} &\equiv 2\xi^{\balpha\bbeta} = \rho^{-1} \omega \left( \partial^{\alpha\beta} V_{\alpha\beta}{}^{\balpha\bbeta} - 5 V_{\alpha\beta}{}^{\balpha\bbeta} \partial^{\alpha\beta} \ln \omega \right) \,, \label{eq:CC10+6}
 \end{split}
\end{equation}
We now see that the outer automorphism of $\SLfo$
\begin{equation}
 V_\alpha{}^{\balpha} \longleftrightarrow \left(V^{-T}\right)_{\balpha}{}^{\alpha} \,, \qquad \partial_{\alpha\beta} \longleftrightarrow \partial^{\alpha\beta} \,, \label{eq:Duality}
\end{equation}
induces a duality on the embedding tensor taking
\begin{equation}
 M_{\balpha\bbeta} \longleftrightarrow \tilde{M}^{\balpha\bbeta} \,, \qquad \tau_{\balpha\bbeta} \longleftrightarrow \tau^{\balpha\bbeta} \,, \qquad \xi^{\balpha\bbeta} \longleftrightarrow \xi_{\balpha\bbeta} \,, \label{eq:DualGaugings}
\end{equation}
where we define $\tau^{\balpha\bbeta} = \frac12 \epsilon^{\balpha\bbeta\bgamma\bdelta} \tau_{\bgamma\bdelta}$ and $\xi_{\balpha\bbeta} = \frac{1}{2} \epsilon_{\balpha\bbeta\bgamma\bdelta} \xi^{\bgamma\bdelta}$.
Thus, given a IIA / IIB truncation gauging only the $\mathbf{10}$'s and $\mathbf{6}$'s there is a consistent truncation of IIB / IIA with the dual gaugings as in \eqref{eq:DualGaugings}. Furthermore, from the Ansatz \eqref{eq:SSAnsatz}, one can check that the reduction formulae for the NS-NS sector remains invariant under this duality. Note that the duality does not mix theories with trombone gaugings and those without.

This structure can also be seen in the half-maximal theory, where the $\mathbf{10} \oplus \mathbf{10}' \oplus \mathbf{6}$ correspond to the usual gaugings while the second $\mathbf{6}$ is the trombone of the half-maximal theory. As first observed in \cite{Dibitetto:2012rk}, the quadratic constraint of the half-maximal theory is invariant under the duality \eqref{eq:DualGaugings}, showing that in the half-maximal theory these gaugings come in pairs of equivalent theories. Here we realise this as a duality between IIA and IIB truncations, which in the maximal theory have inequivalent gaugings.

Let us now discuss more general embedding tensors, where the gaugings are not restricted to lie in the $\mathbf{10}$'s and $\mathbf{6}$'s. Such gaugings can be generated by relaxing the twist Ansatz \eqref{eq:DiagAnsatz}. In general, such gaugings cannot be dualised as discussed in more detail in~\cite{Malek:2015hma}. In particular, gaugings which involve the $\mathbf{4}$'s cannot be dualised. The differential equations governing the gaugings of the $\mathbf{4} \subset \mathbf{15}$ and $\mathbf{4} \subset \mathbf{10}$ are more restrictive than those governing the $\mathbf{4}' \subset \mathbf{40}'$. Thus, while reductions that lead to gaugings in the $\mathbf{4}$ can be dualised to give gaugings in the $\mathbf{4}'$, the reverse is no possible in general. Furthermore, one can check that unlike exchanging the $\mathbf{10}$'s and $\mathbf{6}$'s as in \eqref{eq:DualGaugings}, exchanging $\mathbf{4} \longleftrightarrow \mathbf{4}'$ is not a symmetry of the quadratic constraint \eqref{eq:QC}.

\section{Example: IIA / IIB on $S^3$ and $H^{p,q}$}
Let us revisit the twist matrices of~\cite{Hohm:2014qga} for the consistent truncation of IIA on $S^3$ and $H^{p,q}$. In the notation used here, they correspond exactly to a twist matrix of the form \eqref{eq:DiagAnsatz} and provide gaugings in the $\mathbf{10} \subset \mathbf{15}$. Thus, the duality discussed above can be used to dualise these truncations to obtain the consistent truncation of IIB on the same internal spaces. The resulting seven-dimensional maximal gauged SUGRAs have an embedding tensor in the $\mathbf{10}' \subset \mathbf{40}'$ leading to the gauge group $\mathrm{CSO}(p,q,r) \times \left(U(1)\right)^{4-p-q}$.

\section{A no-go theorem}
So far we have seen that the $\mathbf{10} \subset \mathbf{15}$ and $\mathbf{10}' \subset \mathbf{40}'$ gaugings correspond to $H^{p,q}$ truncations of IIA and IIB, respectively. These are related by the duality \eqref{eq:Duality}. A natural question is whether it is possible to obtain a gauging in the $\mathbf{10} \subset \mathbf{15}$ by a truncation of IIB or equivalently a gauging in the $\mathbf{10}' \subset \mathbf{40}'$ by a truncation of IIA.

To answer this question, let us study the consistency equations \eqref{eq:consistent} again. Assume that we have a IIA gauging, i.e. only $\partial_{m4} \neq 0$. Then we find that in order for a gauging to be uplifted to IIA it must satisfy
\begin{equation}
 W^{\bba\bbb,\bbc} E_{\bba\bbb}{}^{54} = W^{\bba\bbb,\bbc} E_{\bba\bbb}{}^{4m} = 0 \,, \label{eq:NoGoIIA}
\end{equation}
where
\begin{equation}
 W^{\bba\bbb,\bbc} = Z^{\bba\bbb,\bbc} - 3 \epsilon^{\bba\bbb\bbc\bbd\bee} \tau_{\bbd\bee} \,, \quad E_{\bba\bbb}{}^{ab} = \rho^{-1} U_{\bba\bbb}{}^{ab} \,.
\end{equation}
This restricts the possible twist matrices given a certain gauging. In particular let us take $\tilde{M}^{\balpha\bbeta} = \eta^{\balpha\bbeta}$ as required for the $\mathbf{10}' \subset \mathbf{40}'$. If $\eta^{\balpha\bbeta}$ is non-degenerate we find \eqref{eq:NoGoIIA} requires the twist matrix to vanish. Thus, there is no IIA truncation yielding a non-degenerate gauging in the $\mathbf{10}' \subset \mathbf{40}'$ and by the duality there is no IIB truncation with gauging in the $\mathbf{10} \subset \mathbf{15}$. For the degenerate case, the truncation depends on less than three of the internal coordinates so that IIA / IIB coincide.

Let us conclude this section by giving the equivalent requirement for a gauging to be obtained by a IIB truncation. This is given by
\begin{equation}
 W^{\bba\bbb,\bbc} E_{\bba\bbb}{}^{mn} = 0 \,.
\end{equation}

\section{Conclusions}
We showed how to dualise consistent truncations of IIA and IIB supergravity to seven-dimensional maximal gauged SUGRA. The duality is generated by an outer automorphism of $\SLfo \simeq \mathrm{Spin}(3,3)$. Using the results of~\cite{Hohm:2014qga} this allows us to prove the consistent truncation of IIB on $H^{p,q}$ and deduce its non-linear truncation Ans\"atze.

The duality leaves the NS-NS sector invariant and can be used to dualise IIA / IIB truncations with vanishing embedding tensor in the $\mathbf{4}'$ of $\SLfo$. We also gave a necessary requirement for a seven-dimensional gauged SUGRA to be uplifted to IIA / IIB and used this to show that there is no IIA truncation gauging that $\mathbf{10}' \subset \mathbf{40}'$ and by duality no IIB truncation gauging the $\mathbf{10} \subset \mathbf{15}$. This is related to the fact that the half-maximal gaugings mixing these two irreps must violate the section condition~\cite{Lee:2015xga}.

It would be interesting to study this duality in other dimensions. For $d \le 7$ the embedding tensor has only one irreducible component in addition to the trombone. It is then not immediately clear whether the IIA / IIB truncations would define inequivalent lower-dimensional gauged SUGRAs. An interesting starting point would be truncations to three dimensions where there are known to be two inequivalent $\mathrm{SO}(8)$ gaugings believed to be related to $S^7$ truncations of IIA / IIB.

\end{document}